\documentclass[10pt,twocolumn]{article} 
\usepackage{simpleConference}
\usepackage{times}
\usepackage{graphicx}
\usepackage{amssymb}
\usepackage{url,hyperref}
\usepackage[table]{xcolor}
\begin{document}

\title{Multi-Task Multi-Scale Learning For Outcome Prediction in 3D PET Images}

\author{
  Amine Amyar \\
  General Electric Healthcare\\
University of Rouen\\
Rouen, France\\
  \texttt{amyaramine@gmail.com} \and
  Romain Modzelewski \\
  Centre Henri Becquerel \\
University of Rouen\\
Rouen, France \and
   Pierre Vera \\
  Centre Henri Becquerel \\
University of Rouen\\
Rouen, France \and
   Vincent Morard \\
   General Electectric Healthcare \\
   Buc, France \and
   Su Ruan \\
University of Rouen\\
Rouen, France\\ 
}

\maketitle
\thispagestyle{empty}

\begin{abstract}
\textbf{Background and Objectives:} Predicting patient response to treatment and survival in oncology is a prominent way towards precision medicine. To that end, radiomics was proposed as a field of study where images are used instead of invasive methods. The first step in radiomic analysis is the segmentation of the lesion. However, this task is time consuming and can be physician subjective. Automated tools based on supervised deep learning have made great progress to assist physicians. However, they are data hungry, and annotated data remains a major issue in the medical field where only a small subset of annotated images is available. 

\textbf{Methods:} In this work, we propose a multi-task learning framework to predict patient's survival and response. We show that the encoder can leverage multiple tasks to extract meaningful and powerful features that improve radiomics performance. We show also that subsidiary tasks serve as an inductive bias so that the model can better generalize. 

\textbf{Results:}
 Our model was tested and validated for treatment response and survival in lung and esophageal cancers, with an area under the ROC curve of 77\% and 71\% respectively, outperforming single task learning methods. 
 
 \textbf{Conclusions:} We show that, by using a multi-task learning approach, we can boost the performance of radiomic analysis by extracting rich information of intratumoral and peritumoral regions.
\end{abstract}


\section{Introduction}
\begin{figure}

		\centering
		\includegraphics[height=3.5cm, width=8cm]{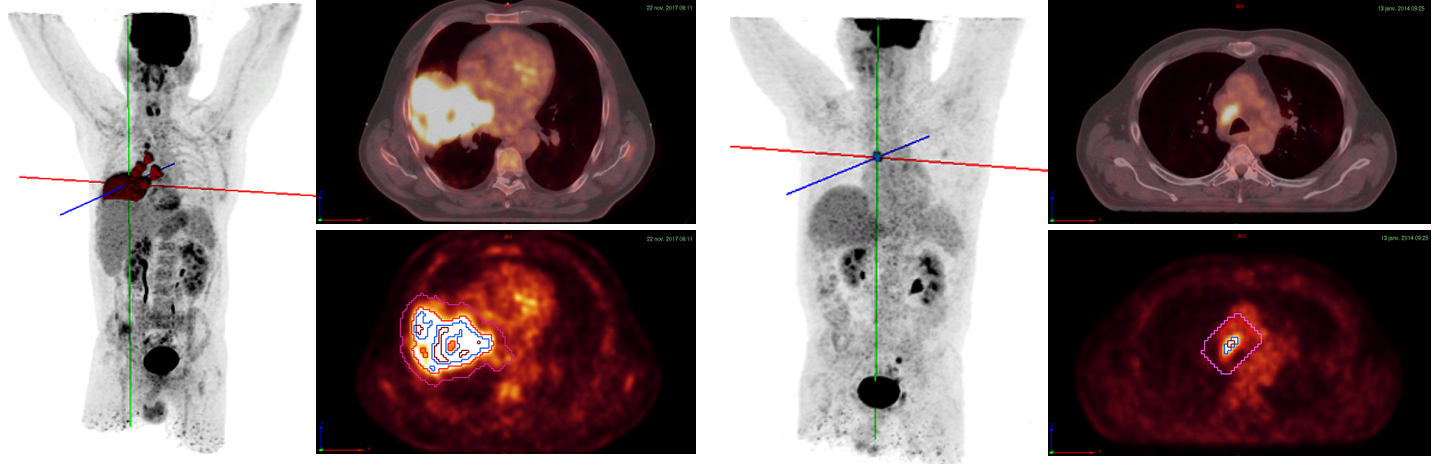}
		\caption{Columns from left to right: Fused PET/CT slice, zoomed on the esophageal tumor seen on FDG-PET only. Metabolic Tumor Volume MTV (40\% SUVmax thresholding) in red. MTV3 (MTV + 3 cm isotropic margin) include the tumor and peritumoral region.}

   \label{fig::pet_cancer}
\end{figure}

Radiomics is a field of study where images have great potential for precision and personalized medicine \cite{lambin2012radiomics,aerts2014decoding}. It is defined as the extraction of a large number of features from medical images such as computed tomography (CT), magnetic resonance imaging (MRI) or positron emission tomography (PET) \cite{kumar2012radiomics}. The first step in radiomic analysis in oncology is the lesion segmentation (Figure \ref{fig::pet_cancer}). This task requires a highly trainable physician, is time consuming and the ground truth defined is physician subjective. Recently, deep learning showed very promising results in image classification \cite{ciregan2012multi}, object detection \cite{szegedy2013deep}, and image segmentation \cite{badrinarayanan2017segnet}. In the medical imaging field, various applications have emerged in different areas, including pathology classification \cite{janowczyk2016deep}, treatment response prediction \cite{amyar2018radiomics}, lesions segmentation \cite{kamnitsas2017efficient} and organs at risk segmentation \cite{trullo2017joint}. Thus, artificial intelligence in general and deep learning in particular can come in handy to develop computer aided diagnostic applications (CAD). However, deep learning approaches are well known for their data hungry nature, and annotated data are usually hard to obtain in the medical imaging field. In a recent work, we tried to tackle this problem with a weakly supervised learning strategy to segment the lesions, and then predict the outcome. This approach showed very promising results outperforming state-of-the-art supervised approaches such as U-Net \cite{ronneberger2015u} for image segmentation, and comparable results with supervised learning \cite{amyar20193} for radiomics analysis. However, the drawback of this method is the two-stage segmentation-outcome prediction. In addition, recent studies have shown the potential of peritumoral regions on boosting the accuracy of outcome prediction \cite{dou2018peritumoral,braman2019association,prasanna2017radiomic}. Thus, the association of the intratumoral and peritumotal regions provides rich information for radiomic analysis \cite{braman2017intratumoral, hu2020assessment}. 

Multi-task learning (MTL) \cite{caruana1997multitask} is a type of learning algorithm that aims to combine several pieces of information from different tasks to improve the model performance and its ability to better generalize \cite{zhang2017survey}. The basic idea of MTL is that different tasks can share a representation of common characteristics \cite{zhang2017survey}, and thus train them jointly. The use of different data sets from different tasks allows learning an efficient representation of the common characteristics of all tasks, because all data sets are used to obtain it, even if each task has a small data set, thus improving the performance of each task.

{\bf Contribution:} In this work, we tackle the challenging problem of training a neural network to classify the pathology, segment the lesion, reconstruct the image, and predict the outcome based on the segmentation results. We believe that the global information in the entity image volume describing the relationship between the tumor and other organs is also useful as are the characteristics of the tumor. We show that, by using a multi-task learning approach, we can boost the performance of radiomic analysis while extracting rich information of intratumoral and peritumoral regions. Our main contributions are summarized as follows:

\begin{enumerate}
    
    \item Our proposed architecture is the first to use jointly global features extracted from entire image and local features from tumor regions to predict the outcome in a radiomics study.
    \\
	\item We design a new multi-tasking learning network to jointly segment the tumor on a 3D PET image and predict the outcome, which is simultaneously associated with two subsidiary tasks, classification and reconstruction. The last two tasks are added to make the features more relevant and also to serve as an inductive bias to better generalize. 
	\\
	\item We utilize a multi-scale feature extraction so that the model can predict the outcome from tumor and tumor neighborhoods features, and also global features at the encoder level.
	\\
	\item We conduct extensive validation strategy with multiple ablation experiments, comparison with state of the art methods in both supervised and multi-task learning.
	
\end{enumerate}

The paper is organized as follows. In Section 2 we review related works on multi-task learning for image classification and segmentation. In Section 3, we describe our multi-task model, which is mainly based on segmentation and outcome prediction tasks. Section 4 presents the experimental studies. In section 5, present the results of our work. Sections 6 and 7 are for discussion and conclusion.

\section{Related Work}

In previous studies, several methods for segmentation of the region of interest and joint classification have been proposed. For instance, Yang et al. \cite{yang2017novel} created a multi-task deep neural network for skin lesion analysis to solve different tasks simultaneously such as lesion segmentation and two independent binary lesion classifications. The MTL model improved learning efficiency and prediction accuracy for each task, in comparison to single task models. They achieved an average Jaccard score of 0.724 for lesion segmentation, while the average values of the area under the receiver operating characteristic curve (AUC) on two lesion classifications are 0.880 and 0.972, respectively. The model consists of a common encoder for the 3 tasks based on GoogleNet \cite{szegedy2015going}, one decoder for segmentation and two fully connected branches for classification. In \cite{asgari2019multiclass} Asgari et al. proposed a multi-class segmentation as multi-task learning for drusen segmentation in retinal optical coherence tomography. The model is based on a multi-decoder architecture that tackles drusen segmentation as a multi-task problem. Instead of training a multi-class model for two classes segmentation, they used one decoder per target class and an extra task for the area between the layers. The model was validated on a dataset of 366 images. They achieved a mean dice of 0.73 compared to 0.68 with multi-class U-Net or 0.66 with a binary U-Net.

In \cite{thome2019multitask} Thome et al. proposed a multi-task classification and segmentation model for cancer diagnosis in mammography. The architecture is based on a fully convolutional network (FCN) \cite{long2015fully}. The model was evaluated on the DDSM database \cite{heath2000pkegelmeyer} with cancer classification  and  pixel  segmentation  with  five  classes. They showed that the model could learn shared representations that are beneficial for both tasks when trained in MTL approach compared to STL. The model achieved a mean dice of 38.28\% and an AUC of 84.02\% compared to a mean dice of 34.98\% and an AUC of 81.37\% for STL. In \cite{he2020multi} He et al. used a multi-task learning approach for the segmentation of organs at risk with label dependence. They used a MTL to accurately determine the contour of organs at risk in CT images. They used an encoder-decoder framework for two tasks. The main task is the segmentation of organs, while the secondary task is the multi-label classification of organs. 
While previous studies showed the advantage of using MTL compared to STL U-Net for image segmentation, recent works have shown the benefit of using U-Net, V-NET \cite{milletari2016v} or Faster-RCNN \cite{ren2016faster} as the backbone network. In \cite{playout2018multitask}, Playout et al. proposed an extension to U-Net architecture relying on multi-task learning with one common encoder, and two decoders to jointly detect and segment red and bright retinal lesions which are essential biomarkers of diabetic retinopathy. At the encoder level, they used residual connections at every scale, mixed pooling for spatial compression and large kernels for convolutions at the lowest scale. Segmentation results are refined with conditional random fields (CRF) and the model is trained with Kappa-based function loss. They achieved a sensitivity of 66,9\% and a specificity of 99,8\% on a public dataset. 

\begin{figure}

		\centering
		\includegraphics[height=6cm, width=9cm]{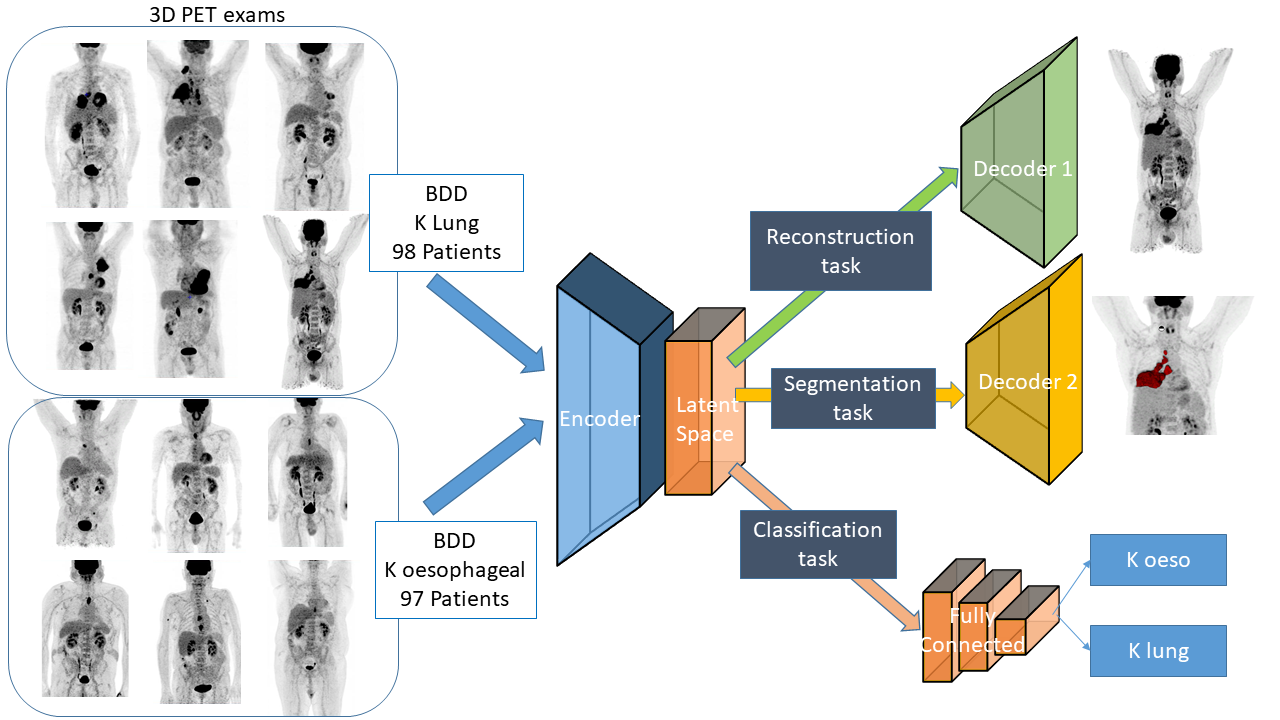}
		\caption{Hard parameter sharing for multi-task learning in deep neural networks used in our proposed architecture.}

  \label{fig::hard_sharing}
\end{figure}

In \cite{vesal2018multi} Vesal et al. proposed a multi-task framework for skin lesion detection and segmentation. The model is based on Faster-RCNN to generate bounding boxes for lesion localization in each image, and "SkinNet" \cite{vesal2018skinnet}, which is a modified version of U-Net. The model was trained and evaluated on ISBI 2017 challenge and the PH2 datasets, outperforming other STL methods in terms of dice coefficient (0.93), Jaccard index (0.88), accuracy (0.96) and sensitivity (0.95), across five-fold cross validation experiments. In \cite{zhou2020multi} Zhou et al. used an MTL framework for segmentation and classification of tumors in 3D automated breast ultrasound images. The main motivation behind their work is the correlation between tumor classification and segmentation, therefore learning these two tasks jointly may improve the outcomes of both tasks. The framework is based on an encoder-decoder network for segmentation and a light-weight multi-scale network for classification, with VNet as the backbone.
These methods cannot be directly applied to 3D PET images to jointly segment the lesion, classify the pathology, and predict the outcome. For instance, the tumor boundaries in PET images for esophageal cancer are not well defined, and sometimes hard to separate from another normal fixation (no tumor). In addition, peritumoral which is defined as the pathology around the tumor is an important information that can boost the prediction accuracy, but it is not considered with previous and classical approaches. Finally, due to the variation in size of the tumors, a multi-scale approach could be a benefice to capture small features as well as investigating bigger ones. In this paper, we take advantage of previous proposed methods and propose a new architecture for radiomics analysis. The main tasks are outcome prediction and lesion segmentation, and the secondary tasks are image reconstruction and pathology classification. We propose a multi-scale feature learning for the outcome prediction, by jointly predicting on the local features and global ones.

\begin{figure*}

		\centering
		\includegraphics[height=11cm, width=15cm]{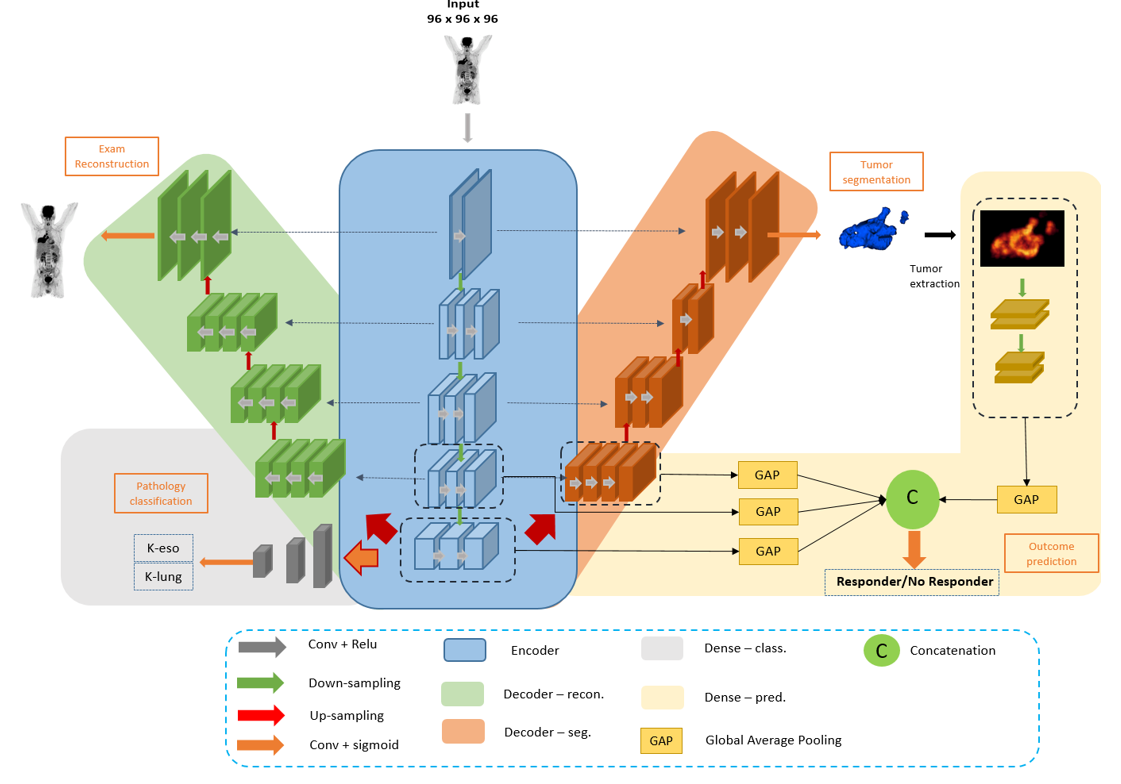}
		\caption{Our proposed architecture, composed of an encoder and two decoders for image reconstruction and tumor segmentation. A fully connected layers are added for classification (esophageal vs lung cancer), and a multi-scale outcome prediction.}

  \label{fig::multitask_architecture}
\end{figure*}

\section{Methods}

Two major strategies are used when training a MTL algorithm, hard parameters sharing \cite{caruana1997multitask} or soft parameters sharing \cite{ruder2017overview}. Hard parameter sharing is the most commonly used approach to MTL in neural networks and greatly reduces the risk of overfitting \cite{ruder2017overview}, see figure \ref{fig::hard_sharing}. In this work we utilize hard parameters sharing due to its great performance and wide utilization.

The reconstruction and pathology classification are extra tasks that serve as an inductive bias. The power of MTL framework lay in the fact that it is able to determine how tasks are related without being given an explicit training signal for task relatedness. 
 
\subsection{Model description}

We propose a new architecture to jointly segment the lesion, classify the pathology, reconstruct the image and predict the outcome. The proposed network is shown in figure \ref{fig::multitask_architecture}. We use U-Net as the backbone due to its great performance in 3D medical image segmentation. The architecture consists of four parts: (i) a common encoding part, (ii) a decoding part for reconstruction, iii) a decoding part for the segmentation and (iii) skip connections between them, which form a W, see figure \ref{fig::multitask_architecture}. To that we add a multi-layer perceptron (MLP) for the classification task, and a convolutional neural network (CNN) for the outcome prediction based on the segmentation result. Finally, we use multi-scale approach to feed global features to the CNN, to make a prediction on both global features and tumor ones. To summarize, many classic image classification networks use transfer learning \cite{pan2009survey} to extract high level features from CNN models, such as VGG16 \cite{simonyan2014very} or ResNET \cite{he2016deep}. Motivated by this, we use the same encoder for lesion segmentation and pathology classification to extract common features. We add reconstruction task as a secondary task so that the neural network can extract meaningful features about PET images. 

\subsubsection{Encoder-Decoder}

 The encoder is used to obtain the disentangled feature representation. It is a 10-layers 3D convolutional neural networks with convolution filters of $3\times3\times3$ and a maxpooling of $2\times2\times2$ after each 2 convolutional layers and a skip connection. The number of feature maps increases from 64 for the 2 first layers to 1024 for the last ones. We use $relu$ activation function and a Dropout of $0.5$ after the last convolutional layer. The structure of the 2 decoders is the same, with upsampling to return to the original image size followed by convolutional layers to reduce the number of features by a factor of 2. These features are concatenated with the ones from the corresponding level of the encoder.

\subsubsection{Multi-scale Feature Extraction}

\begin{figure*}

		\centering
		\includegraphics[height=10cm, width=15cm]{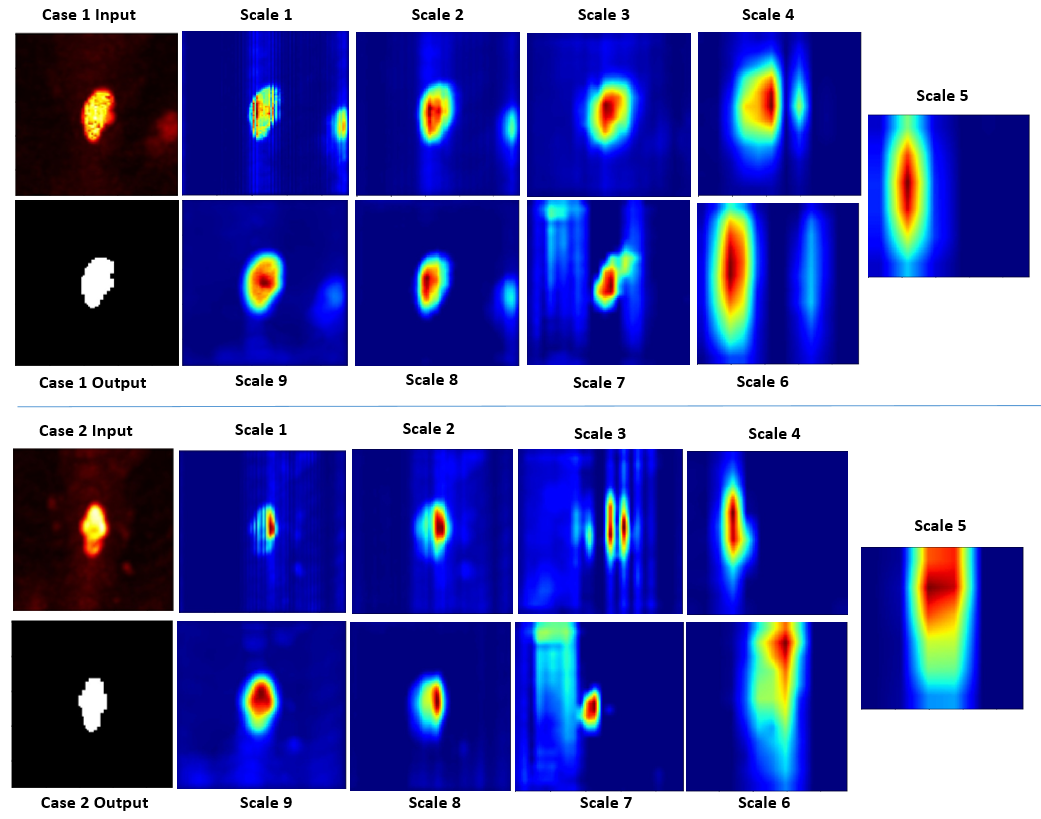}
		\caption{Heatmap generated using grad-cam. The input is a cropped image around the lesion and the output is the mask. The neural network focuses on additional information other than the lesion at scale 4, 5 and 6.}

  \label{fig::multi_scale}
\end{figure*}

For the outcome prediction, we take advantage of both local features and global features. Local features are extracted from the segmentation result, while the global features are obtained from the common encoder. To benefit from features of different scales, we designed a multi-scale feature concatenation model for the radiomics task, as shown in figure \ref{fig::multi_scale}. We concatenate feature maps from Level 3 to 5 in the encoder with the convolutional network in the outcome prediction. As a strong tool to evaluate and analyze the decision made by the neural network, we visualize heatmaps at different levels of the encoder and the decoder for the segmentation. To visualize the heatmaps, we use Grad-cam technique \cite{selvaraju2017grad} to produce visual explanation at each scale. We can observe that scale 4, 5 and 6 extract rich features at the tumor level and beyond, including peritumotal regions and other important fixations. To incorporate this information at the prediction level we design a multi-scale feature concatenation model by fusing feature maps from scale 4, 5 and 6 with the tumor features. We use a channel-wise global average pooling (GAP) as in \cite{zhou2020multi} to reduce the complexity in training time and to keep also important features, since it is more robust to spatial translation.

\makeatletter
\newcommand{\thickhline}{%
    \noalign {\ifnum 0=`}\fi \hrule height 2pt
    \futurelet \reserved@a \@xhline
}

\makeatother

\subsubsection{The reconstruction task T1}

 We trained the model with a linear activation for the output and a mean squared error for the loss function (Lrecon) and used accuracy as the metric:
 \begin{equation}
     Lrecon = \displaystyle\frac{1}{n}\sum_{t=1}^{n}(y\_true - y\_predict)^2
 \end{equation}
where y\_true is the true label and y\_predict is the predicted label.

\subsubsection{The segmentation task T2}

The decoder for the segmentation is similar to the one used for the image reconstruction with the exception of the last activation function. We substituted the mean squared error by a sigmoid for a binary output (tumor/no tumor). The loss function is the dice coefficient loss (Lseg):
 \begin{equation}
     dice\_coef = \displaystyle\frac{2 * |X \cap Y| + \epsilon}{|X| + |Y| + \epsilon}
 \end{equation}
 \begin{equation}
     Lseg = - dice\_coef
 \end{equation}
where $\epsilon$ is the the smoothing factor and used to avoid a division by zero. 

\subsubsection{The classification task T3} 
The resulting set of feature maps encloses the entire spatial local information, as well as the hierarchical representation of the input. The activation of the i\textsuperscript{(th)} unit of the h hidden layer is given by:
\begin{equation}
	h\textsubscript{i} = f(b\textsubscript{i} + Wh\textsubscript{i} * V) \qquad \textnormal{with } i = 1,.....,H.
	\label{eq::conv}
\end{equation}

In details, the output of the encoder is a tensor of mini\_batch x 32 x 32 x 1024 to which we add a convolutional layer followed by a maxpooling followed by a flatten. The multilayer perceptron consists of a two Dense layer with 128 and 64 neurons respectively, with a dropout of 0.5 and the activation function $elu$. The loss function is a binary cross entropy:
\begin{equation}
    Lclass = -\frac{1}{n} \sum_{i=1}^{n}[y_{i}\log(\hat{y}_{i}) + (1-y_{i})\log(1-\hat{y}_{i})]\\
	\label{eq::binary cross-entropy}
\end{equation}

which is a special case of the multinomial cross-entropy loss function for m = 2 :
\begin{equation}
L(\theta) = -\frac{1}{n}\sum_{i=1}^{n}\sum_{j=1}^{m}y_{ij}\log(\hat{y}_{ij}) 
\end{equation}
where n is the number of patients and y is the class label (esophageal cancer, lung cancer). 

\subsubsection{The prediction task T4} 
The prediction branch is connected to three layers from the encoder and segmentation decoder to incorporate global features, in addition to tumor features extracted from the segmentation result. It is composed of 2 convolutional layers with 64 feature maps each followed by a max-pooling and 2 other convolutional layers with 128 feature maps each. Then, we apply a global average pooling to concatenate tumor-based features (local features) with encoder-decoder global features in a multi-scale. Finally, three fully connected layers are used for the prediction with 128, 128 and 1 neurons respectively. The loss function is the binary cross-entropy, and the performance metric is the accuracy:
\begin{equation}
    Lpredict = -\frac{1}{n} \sum_{i=1}^{n}[y2_{i}\log(\hat{y}2_{i}) + (1-y2_{i})\log(1-\hat{y}2_{i})]\\
	\label{eq::binary cross-entropy}
\end{equation}

where n is the number of patients and y is the outcome.

\subsubsection{multi-task Loss Function}


We use reconstruction task to learn more meaningful features of PET exams, and outcome prediction task so that the network will focus attention on the most discriminator regions for the segmentation task so that. In our experiments, the Adam optimizer \cite{kingma2014adam} algorithm was used with a mini batch of 4 and a learning rate of 0.0001. The global loss function (loss glob) for the 4 tasks is defined by:

\begin{equation}
    \bf{loss\_glob} = \alpha Lrecon + \beta Lseg + \omega Lclass + \lambda predict
\end{equation}

where $\alpha$ = $\beta$ and $\omega$= 1-($\lambda  + 2 \times \alpha$). Our model was trained for 1500 epochs with an early stopping of 70.

\subsection{Dataset}
Our experiments were run on 195 PET image volumes with lung (98) and esophageal (97) cancer, from Henri Becquerel Center, Rouen, France. All patients underwent whole body FDG PET with a CT (baseline PET), at the initial stage of the pathology and before any treatment. The reconstructed exam voxel size was $4.06\times4.06\times2.0$ mm\textsuperscript{3} and were spatially normalized by re-sampling all the dataset to an isotropic resolution of $2\times2\times2$ mm\textsuperscript{3} using the k-nearest neighbor interpolation algorithm. We split the data into 2 groups to train and test the deep learning methods. One group was used for training the models (77 esophageal and 78 Lung) and one locked group for testing (40 patients). Furthermore, for the CNN, the training samples were split into 2 groups, a train set (57 esophageal and 58 Lung) and a validation set (40 patients).

\makeatother

\subsection{Implementation}
All models were implemented using python and keras deep learning library, with tensorflow as backend, and trained on nvidia p6000 quadro gpu with 24gb. Some tested state-of-the-art models were developed using pytorch library.
\section{Experiments}
We compare the performance of STL and MTL models. We present an empirical test that rules out these mechanisms and thus ensures that the benefit from MTL is due to the information in the extra tasks.

\textbf{Experiment 1}: The first experiment consists of the optimization of the network by testing the different combination of tasks. The models developed include single tasks models, 2 and 3 tasks models, and all tasks’ models. Reconstruction and pathology classification are secondary tasks; thus, they are combined either with segmentation or outcome prediction or both of them. Also, outcome prediction with and without global features were evaluated, in addition to outcome prediction with and without local features.  In total, 15 models were developed for the outcome prediction. 
\textbf{Experiment 2:} The second experiment is to evaluate the performance of the best model with state of the art methods for image segmentation such as U-Net, V-NET and a weakly supervised multi-task approach \cite{amyar2020weakly}. The WSL-MTL model uses a priori knowledge by defining two points in two 2D maximum intensity projection (MIP) images for coronal and sagittal views. The model learns to classify the two MIPs into lung and esophageal cancers, and by generating a class activation map (CAM), it calculates a distance between the CAM generated and the two points defined and learn to minimize the distance between the CAMs and the 2 points in a multi-task learning approach. Finally, the corrected CAMs for sagittal and coronal views are used to retrieve the tumor in the 3D space.

\textbf{Experiment 3:} The third experiment is to compare our models with state of the art methods for image classification and outcome prediction. We use: Alexnet \cite{krizhevsky2017imagenet}, VGG-16 \cite{simonyan2014very}, ResNET50 \cite{he2016deep}, 169-layer DenseNet \cite{huang2017densely} and InceptionV3 \cite{szegedy2016rethinking}. We compare also our results with deep radiomics such as 3D RPET-NET \cite{amyar20193} and a 6 layers 3D convolutional neural network. 

\textbf{Experiment 4:} In experiment 4 we study the effects of $\lambda$ on the multitask learning. We have tested different values: 0.1, 0.3, 0.5, 0.7 and 0.9.

\textbf{Experiment 5:} Finally, we compare our proposed method with state of the art multi-task methods,  including \cite{zhou2021multi},\cite{wang2018simultaneous}, \cite{qu2019joint} and \cite{chen2018multi}. We extended 2D networks to 3D. To incorporate both local and larger contextual information, we employ a multi-scale feature extraction for outcome prediction.

\section{Results}

The performances of the models were evaluated using the dice coefficient for the segmentation task, and the area under the ROC curve (AUC) and the accuracy (Acc), for both classification and prediction.
\definecolor{Gray}{gray}{0.85}
\definecolor{LightCyan}{rgb}{1,1,0.6}
\definecolor{LightRed}{rgb}{0.9,0.7,0.5}
\newcolumntype{a}{>{\columncolor{Gray}}c}
\newcolumntype{b}{>{\columncolor{LightCyan}}c}
\newcolumntype{d}{>{\columncolor{LightRed}}c}

\begin{table*}[!ht]
\centering
\begin{tabular}{lllldaabb}
\thickhline
& & & & \bf Seg.& \bf Class. &  & \bf Pred. &
\\
\thickhline
\bf  &\bf  Tasks  & \bf   Global features  & \bf    Tumor features & \bf Dice\_coef & \bf Accuracy & \bf AUC & \bf Accuracy & \bf AUC \\
\hline

Esoph. & T1 \& T4 & \checkmark & x & / &  / & / & 0.60 & 0.60 \\

&T1 \& T3 \& T4  & \checkmark & x & / & 0.95  & 0.94 & 0.65 & 0.68  \\

&T1 \& T2 \& T4 & \checkmark & x & 0.77 & / & / &  0.65 &  0.64  \\

&T1 \& T2 \& T4 &  x & \checkmark & 0.76 &  / & / & 0.65 &  0.63 \\

&T1 \& T2 \& T4 &  \checkmark & \checkmark & 0.74 & / & / & 0.70 & 0.63 \\

&T2 \& T4 &  \checkmark & x & 0.73 & / & / & 0.70  & 0.71  \\

&T2 \& T4 &  x & \checkmark & 0.70 & / & / & 0.70  & 0.74 \\

&T2 \& T4 &  \checkmark & \checkmark & \bf{0.79}  & / & / & 0.73 & 0.72 \\

&T2 \& T3 \& T4 &  \checkmark & x & 0.71 &  0.94 & 0.93  & 0.72 & 0.70 \\

&T2 \& T3 \& T4 &  x & \checkmark & 0.69 &  0.91 & 0.92 & 0.70 &  0.71 \\

&T2 \& T3 \& T4 &  \checkmark & \checkmark  & 0.71 & 0.93  & 0.91  & 0.75 & 0.73 \\

&T3 \& T4  &  x & x & / & \bf{0.98} & \bf{0.97} & 0.60 & 0.59 \\

&T1 \& T2 \& T3 \& T4 &  \checkmark & x & 0.73 & 0.96 & 0.95 & 0.70 & 0.67 \\

&T1 \& T2 \& T3 \& T4 & x & \checkmark & 0.75 & 0.97 & 0.95 & 0.76 & 0.74 \\
    
&\bf{T1 \& T2 \& T3 \& T4} & \bf{\checkmark} & \bf{\checkmark} & 0.73 & 0.97& 0.94 & \bf{0.79} &  \bf{0.77}  \\

\thickhline

Lung & T1 \& T4 & \checkmark & x  & / & /  & / &  0.49 &  0.51 \\

&T1 \& T3 \& T4  & \checkmark & x  & / & 0.95 & 0.94 & 0.59 & 0.56 \\

&T1 \& T2 \& T4 & \checkmark & x  & 0.84 &  / & / & 0.60 & 0.58 \\

&T1 \& T2 \& T4 &  x & \checkmark  & 0.84 &  / & / & 0.64 & 0.60 \\

&T1 \& T2 \& T4 &  \checkmark & \checkmark  & 0.83 & / & / & 0.65 & 0.62 \\

&T2 \& T4 &  \checkmark & x  & \bf{0.86}  & / & / & 0.65  &  0.57 \\

&T2 \& T4 &  x & \checkmark  & 0.81 & / & / & 0.67 & 0.65 \\

&T2 \& T4 &  \checkmark & \checkmark & 0.82  & / & / & 0.67 &  0.66 \\

&T2 \& T3 \& T4 &  \checkmark & x  & 0.80 & 0.94 & 0.93 &  0.68 & 0.66 \\

&T2 \& T3 \& T4 &  x & \checkmark   &  0.76  & 0.91 & 0.92 &  0.68 &  0.65 \\

&T2 \& T3 \& T4 &  \checkmark & \checkmark  & 0.79  & 0.93 & 0.91 & 0.69 &   0.67 \\

&T3 \& T4  &  x & x  & / & \bf{0.98} & \bf{0.97} & 0.70 & 0.65  \\

&T1 \& T2 \& T3 \& T4 &  \checkmark & x  & 0.81 & 0.96 & 0.95 & 0.70 &  0.62 \\

&T1 \& T2 \& T3 \& T4 & x & \checkmark  &  0.83 & 0.97 & 0.95 & \bf{0.71} & 0.69 \\

&\bf{T1 \& T2 \& T3 \& T4} & \bf{\checkmark} & \bf{\checkmark} & 0.82 & 0.97 & 0.94 & 0.70 &  \bf{0.71}  \\

\thickhline

\end{tabular}
\caption{Results of experiment 1: segmentation, classification and prediction results from different scenarios, for esophageal and lung cancers. T1: reconstruction, T2: segmentation, T3: pathology classification, T4: outcome prediction.}

\label{tab::mtl_experiment1}

\end{table*}

The main results of the five experiments are shown in Tables 1 to 5. The neural network was trained for 1500 epochs with an early stopping of 70. 

\begin{table}
\centering
\begin{tabular}{lld}
\thickhline
             &  Method                         &  Dice coef   \\ \hline

\hline

Esophageal cancer & U-NET (T2) & 0.69 \\
& V-NET & 0.69 \\
& WSL-MTL & 0.73 \\
&\textbf{Ours} & \bf{0.73} \\
\hline
Lung cancer & U-NET (T2) & 0.80 \\
&  V-NET & \bf{0.85} \\
& WSL-MTL  & 0.77 \\
&\textbf{Ours} & 0.82 \\
\thickhline

\end{tabular}
\caption{Experiment 2: Segmentation results for esophageal and lung cancer compared to the state of the art methods. WSL: weakly supervised learning model developed in \cite{amyar2020weakly}.}

\label{tab::mtl_experiment2}

\end{table}

\begin{table*}
\centering
\begin{tabular}{llaabb}

\thickhline
             &                           &  \bf Classification          &            &  \bf Prediction          &     \\ 
\hline
\thickhline
             &  Method                         &  Accuracy          &  AUC          &  Accuracy          &  AUC      \\ 
\hline

Esophageal cancer & AlexNet & 0.74 & 0.73 & 0.54 & 0.52 \\
& VGG-16 & 0.79 & 0.77 & 0.53 & 0.51 \\
& VGG-19 & 0.78 & 0.78 & 0.55 & 0.53 \\
& ResNet50 & \bf{0.97} & \bf{0.97} & 0.62 & 0.63 \\
& 169-layers DenseNet & 0.95 & 0.94 & 0.63 & 0.61 \\
& InceptionV3 & 0.93 & 0.92 & 0.61 & 0.69 \\
& 3d-rpet-net  & / & / & 0.72 & 0.70 \\
& WSL-MTL  & / & / & 0.69 & 0.67 \\
& 6 layers CNN & 0.80 & 0.81 & 0.69 & 0.68  \\
& Ours & \bf{0.97} & 0.94 & \bf{0.79} &  \bf{0.77}  \\
\thickhline

Lung cancer & AlexNet & 0.74 & 0.73 & 0.51 & 0.49 \\
& VGG-16 & 0.79 & 0.77 & 0.50 & 0.51 \\
& VGG-19 & 0.78 & 0.78 & 0.51 & 0.51 \\
& ResNet50 & \bf{0.97} & \bf{0.97} & 0.59 & 0.57 \\
& 169-layers DenseNet & 0.95 & 0.94 & 0.61 & 0.60 \\
& InceptionV3 & 0.93 & 0.92 & 0.57 & 0.59 \\
& 3d-rpet-net  & / & / & 0.68 & 0.61 \\
& WSL-MTL  & / & / & 0.65 & 0.59 \\
& 6 layers CNN & 0.80 & 0.81 & 0.63 & 0.60  \\
& Ours & \bf{0.97} & 0.94 & \bf{0.70} &  \bf{0.71} \\

\thickhline
\end{tabular}
\caption{Experiment 3: classification and outcome prediction results compared to state of the art methods for esophageal and lung cancers.}

\label{tab::mtl_experiment3}
\end{table*}

\textbf{Experiment 1}: As shown in Table \ref{tab::mtl_experiment1}, the best results for outcome prediction were obtained with the combination of the four tasks with multi-scale, and with tumor and global features. It achieved an accuracy of 0.79 and AUC of 0.77 for esophageal cancer outperforming 14 other scenarios which are composed of several combination of different tasks, with and without multi-scale and with and without tumor features. For lung cancer, our proposed model achieved an accuracy of 0.70 and AUC of 0.71 in multi-scale, and an accuracy of 0.71 and AUC of 0.69 when using only tumor features. Using only reconstruction and prediction resulted in a poor performance for both lung and esophageal cancers. For the segmentation, the best results were achieved by the combination of segmentation and prediction in multi-scale for esophageal cancer (dice coefficient = 0.79) and using only global features for lung cancer (dice coefficient = 0.86). Our proposed model achieved a dice score of 0.73 in multi-scale and 0.75 when using only tumor features for esophageal cancer, and a dice score of 0.82 in multi-scale and 0.83 when using only tumor features for lung cancer. The combination of the reconstruction, segmentation and prediction also resulted in good results for segmentation, when using only global features, only tumor features and in multi-scale: 0.77, 0.76 and 0.74 for esophageal cancer and 0.84, 0.84, 0.83 for lung cancer respectively. When using the classification task in addition to segmentation and prediction the performance on segmentation decreases: 0.71, 0.69 and 0.71 for esophageal cancer and 0.80, 0.76, 0.79 for lung cancer. This can be explained due to the fact that the reconstruction task helps in the extraction of rich meaningful features that contribute to the segmentation better than the classification task. For the classification, the best results were achieved with the combination of the classification and prediction tasks, without reconstruction and segmentation: accuracy = 0.98 and AUC = 0.97. Our proposed model achieved an accuracy of 0.97 and an AUC of 0.94 and 0.95 with multi-scale and only tumor features respectively. Since the goal of our study is to focus on the prediction task, the performance of the other 2 tasks (segmentation and classification) can be a little higher when using only segmentation and prediction or classification and prediction, but not for the prediction. This is because to improve the performance of the prediction task, the model tends to find the most informative and discriminating region in the image that allows this improvement. This results in the extraction of intratumoral and peritumoral tumor regions, which may differ from segmentation ground truth but improve the prediction. The combination of segmentation and prediction resulted in an accuracy of 0.70, 0.70 and 0.73 and an AUC of 0.71, 0.74 and 0.72 for esophageal cancer for global features, tumor features and multi-scale respectively, and an accuracy of 0.65, 0.65 and 0.67 and an AUC of 0.57, 0.65 and 0.66 for lung cancer. The combination of the classification and prediction resulted in an accuracy of 0.60 and an AUC of 0.59 for esophageal cancer and an accuracy of 0.70 and an AUC of 0.65 for lung cancer.

\textbf{Experiment 2}: In Table \ref{tab::mtl_experiment2}, segmentation results for three other state of the art methods are reported and compared to our proposed model, for esophageal and lung cancers. The 3 models are: U-Net, which represents the task T2 for the segmentation since it was used as the backbone in our model, V-Net and a weakly supervised multi-task learning (WSL-MTL) model for tumor segmentation. Our model achieved the best results with the WSL-MTL for esophageal cancer (dice coefficient = 0.73), and slightly worse than the WSL-MTL for lung cancer (dice coefficient = 0.82 and 0.85 respectively), since the WSL-MTL was trained to do the segmentation as a primary objective. Our model was better than single task (T2) U-Net and V-Net:  0.69 and 0.69 for esophageal cancer and 0.80 and 0.77 for lung cancer. These results show that our model can correctly find the tumor regions from which local tumor features can well be extracted.
\textbf{Experiment 3}: Table \ref{tab::mtl_experiment3} shows the results of the third experiment. We compared our method with state-of-the-art deep learning models for image classification and prediction. Our proposed model outperformed other methods for the prediction task for both esophageal and lung cancers. ResNet50 had slightly better results for the classification (accuracy 0.97 and AUC 0.97) but very poor results the prediction: accuracy = 0.62 and AUC = 0.63 for esophageal cancer and accuracy = 0.59 and AUC = 0.57 for lung cancer. AlexNet, VGG-16 and VGG-19 have not shown promising results.

\begin{table*}
\centering
\begin{tabular}{lldaabb}
\thickhline
 & $\lambda$ &  Dice coef   &  Accuracy          &  AUC          &  Accuracy          &  AUC      \\ 
\hline 
Esophageal cancer & 0.1 & \bf{0.74} & 0.94 & 0.94 & 0.70 & 0.69 \\
 & 0.3 &  0.73 & \bf{0.97} & \bf{0.94} & \bf{0.79} &  \bf{0.77}   \\
  & 0.5 & 0.71 & 0.95 & 0.94 & 0.77 & 0.76 \\
   & 0.7 & 0.71 & 0.93 & 0.94 & 0.70 & 0.71 \\
     & 0.9 & 0.69 & 0.89 & 0.88 & 0.73 & 0.71 \\
\thickhline

Lung cancer & 0.1 & 0.80 & 0.94 & 0.94 & 0.65 & 0.64 \\
 & 0.3 & 0.82 & \bf{0.97} & \bf{0.94} &  0.70 &  \bf{0.71} \\
  & 0.5 & \bf{0.85} & 0.95 & 0.94 & \bf{0.71} & 0.70 \\
   & 0.7 & 0.80 & 0.93 & 0.94 & 0.69 & 0.70 \\
      & 0.9 & 0.78 & 0.89 & 0.88 & 0.71 & 0.70 \\

\thickhline
\end{tabular}
\caption{Experiment 4: The effects of $\lambda$ on the multi-task learning.}

\label{tab::mtl_experiment4}
\end{table*}

\begin{table*}
\centering
\begin{tabular}{lldaabb}
\thickhline
 & Method &  Dice coef   &  Accuracy          &  AUC          &  Accuracy          &  AUC      \\ 
\hline 
Esophageal cancer & \cite{zhou2021multi} & \bf{0.75} & 0.96 & 0.94 & 0.73 & 0.71 \\
 & \cite{chen2018multi} & 0.68 & 0.91 & 0.90 & 0.68 & 0.66 \\
  & \cite{qu2019joint} & 0.69 & 0.92 & 0.90 & 0.70 & 0.69 \\
   & \bf{Ours} & 0.73 & \bf{0.97} & \bf{0.94 }& \bf{0.79} &  \bf{0.77} 
    \\
\thickhline

Lung cancer & \cite{zhou2021multi} & 0.77 & 0.96 & 0.94 & 0.67 & 0.65 \\
 & \cite{chen2018multi} & 0.73 & 0.91 & 0.90 & 0.63 & 0.60 \\
  & \cite{qu2019joint} & 0.80 & 0.92 & 0.90 & 0.68 & 0.65 \\
   &    \bf{Ours} & \bf{0.82} & \bf{0.97} & \bf{0.94 }& \bf{0.70} &  \bf{0.71} \\

\thickhline
\end{tabular}
\caption{Experiment 5: A quantitative comparison between our model and state of the art multi-task methods.}

\label{tab::mtl_experiment5}
\end{table*}

\textbf{Experiment 4}: In Table \ref{tab::mtl_experiment4} the the influence of $\lambda$ on the performance of our model is reported. We achieved the best results for tumor classification and outcome prediction with $\lambda = 0.3$. When lowering the value of $\lambda$ the model achieves slightly better result for the segmentation for esophageal cancer (0.74) but a worse prediction result for both esophageal and lung (accuracy = 0.70 and AUC = 0.69, accuracy = 0.65 and AUC = 0.64). For $\lambda = 0.5$ our model achieves comparable results for both pathologies: accuracy = 0.77 and AUC = 0.76 for esophageal cancer and accuracy = 0.71 and AUC = 0.70 for lung cancer, with a better dice coefficient (0.85) for the segmentation of lung tumors. Increasing $\lambda$ does not result in an improvement of the prediction task, it decreases the performance of the segmentation and classification tasks and the prediction: accuracy = 0.73, AUC = 0.71 for esophageal and accuracy = 0.71 and AUC = 0.70 for lung for prediction, accuracy = 0.89 and AUC = 0.88 for classification, and a dice coefficient = 0.69 and 0.78 for esophageal and lung cancers respectively for segmentation.

\textbf{Experiment 5}: Table \ref{tab::mtl_experiment5} reports the results of three state-of-the-art methods for multi-task learning for segmentation and classification. Our proposed model achieves the best results for both esophageal and lung cancers for the prediction and classification task, where \cite{zhou2021multi} achieves a slight improvement on the segmentation task for esophageal cancer (dice coefficient = 0.75), and comparable results for the classification (accuracy = 0.96 and AUC = 0.94). 

\section{Discussion}
We have developed a new deep learning multi-task model to jointly identify esophageal and lung tumors, segment the tumor regions of interest and predict patient's outcome. Our architecture is general, which means that it can be used for other segmentation-classification-prediction applications. We have also compared our method with several state-of-the-art algorithms such as U-NET, V-NET and WSL-MTL for tumor segmentation, methods for image classification and prediction, and for multi-task learning such as \cite{zhou2021multi,qu2019joint,chen2018multi}. To show the performance of our method, we tested the different combinations of different tasks, as well as using only global features or only tumor features and a multi-scale regrouping tumor and global features. We have added the reconstruction task to leverage useful information contained in multiple related tasks to improve both segmentation and prediction performances.  

Multi-task learning can handle small data problems well, although each task can have a relatively small data set. In contrast to conventional radiomics, where only one pathology is studied at a time, multi-task learning allows to study different cancer types at the same time, thus, to increase the size of the dataset and help the model to learn meaningful features from PET images so that help to improve the prediction.

We have added global image features through a multi-scale by using a global average pooling and then concatenated with tumor features to predict the outcome. Having both global and local features help to improve the performance of the model compared to using only tumor features as in classical radiomic. Although the segmentation performance drops a little when combining the 4 tasks compared to segmentation-prediction alone, the most important task in our study is the prediction, hence we let the model decides which is the most important region in the image that increases the prediction performance, resulting in encompassing intratumoral and peritumoral regions. Since dice coefficient measures the intersection between the ground truth and the segmentation result, it can drop a little its score. The segmented tumor region may not be exactly the same as the ground truth, but it may be more relevant for prediction. In our study, the dice coefficient is used to ensure that the result of the segmentation is anatomically correct, not to be perfect. 

One of the main advantages of our proposed method relies in the fact that once the learning is finished, we no longer need segmentation ground truth to do radiomics. The model requires only the PET images as input, thus, to avoid the tedious segmentation task for physicians. Also, the architecture is general. The model can be modified easily to add other cancer types to do radiomics without changing the architecture, just the classification branch.

\section{Conclusion}

In this paper, we proposed a multi-task learning approach to predict patient’s outcome from PET images and segment the regions of interest simultaneously. Our method can improve prediction results even if we have only several small datasets. thanks to learning tasks in parallel while using a shared representation. Therefore, what is learned for each task can help other tasks be learned better. We show also that subsidiary tasks serve as an inductive bias so that the model can generalize better. Our model was tested and evaluated for treatment response and survival in lung and esophageal cancers, outperforming single task learning methods and state-of-the-art multi-task learning methods. In the future, we will add other cancers to validate our framework and develop an attention mechanism to combine the different features.

\section*{Declaration of Competing Interest}
There are no conflicts of interest to disclose for publication of this paper.

\section*{Statements of ethical approval}
The study was approved by the Ethics Committee of The Centre Henri Becquerel. All the participants provided their written informed consent to participate in this study.

\section*{Acknowledgment}
This research has received no external funding.

\bibliographystyle{abbrv}
\bibliography{main}

\begin{thebibliography}{10}

\bibitem{aerts2014decoding}
H.~J. Aerts, E.~R. Velazquez, R.~T. Leijenaar, C.~Parmar, P.~Grossmann,
  S.~Carvalho, J.~Bussink, R.~Monshouwer, B.~Haibe-Kains, D.~Rietveld, et~al.
\newblock Decoding tumour phenotype by noninvasive imaging using a quantitative
  radiomics approach.
\newblock {\em Nature communications}, 5(1):1--9, 2014.

\bibitem{amyar2020weakly}
A.~Amyar, R.~Modzelewski, P.~Vera, V.~Morard, and S.~Ruan.
\newblock Weakly supervised pet tumor detection usingclass response.
\newblock {\em arXiv preprint arXiv:2003.08337}, 2020.

\bibitem{amyar20193}
A.~Amyar, S.~Ruan, I.~Gardin, C.~Chatelain, P.~Decazes, and R.~Modzelewski.
\newblock 3-d rpet-net: development of a 3-d pet imaging convolutional neural
  network for radiomics analysis and outcome prediction.
\newblock {\em IEEE Transactions on Radiation and Plasma Medical Sciences},
  3(2):225--231, 2019.

\bibitem{amyar2018radiomics}
A.~Amyar, S.~Ruan, I.~Gardin, R.~Herault, C.~Clement, P.~Decazes, and
  R.~Modzelewski.
\newblock Radiomics-net: Convolutional neural networks on fdg pet images for
  predicting cancer treatment response.
\newblock {\em Journal of Nuclear Medicine}, 59(supplement 1):324--324, 2018.

\bibitem{asgari2019multiclass}
R.~Asgari, J.~I. Orlando, S.~Waldstein, F.~Schlanitz, M.~Baratsits,
  U.~Schmidt-Erfurth, and H.~Bogunovi{\'c}.
\newblock Multiclass segmentation as multitask learning for drusen segmentation
  in retinal optical coherence tomography.
\newblock In {\em International Conference on Medical Image Computing and
  Computer-Assisted Intervention}, pages 192--200. Springer, 2019.

\bibitem{badrinarayanan2017segnet}
V.~Badrinarayanan, A.~Kendall, and R.~Cipolla.
\newblock Segnet: A deep convolutional encoder-decoder architecture for image
  segmentation.
\newblock {\em IEEE transactions on pattern analysis and machine intelligence},
  39(12):2481--2495, 2017.

\bibitem{braman2019association}
N.~Braman, P.~Prasanna, J.~Whitney, S.~Singh, N.~Beig, M.~Etesami, D.~D. Bates,
  K.~Gallagher, B.~N. Bloch, M.~Vulchi, et~al.
\newblock Association of peritumoral radiomics with tumor biology and
  pathologic response to preoperative targeted therapy for her2
  (erbb2)--positive breast cancer.
\newblock {\em JAMA network open}, 2(4):e192561--e192561, 2019.

\bibitem{braman2017intratumoral}
N.~M. Braman, M.~Etesami, P.~Prasanna, C.~Dubchuk, H.~Gilmore, P.~Tiwari,
  D.~Plecha, and A.~Madabhushi.
\newblock Intratumoral and peritumoral radiomics for the pretreatment
  prediction of pathological complete response to neoadjuvant chemotherapy
  based on breast dce-mri.
\newblock {\em Breast Cancer Research}, 19(1):1--14, 2017.

\bibitem{caruana1997multitask}
R.~Caruana.
\newblock Multitask learning.
\newblock {\em Machine learning}, 28(1):41--75, 1997.

\bibitem{chen2018multi}
C.~Chen, W.~Bai, and D.~Rueckert.
\newblock Multi-task learning for left atrial segmentation on ge-mri.
\newblock In {\em International workshop on statistical atlases and
  computational models of the heart}, pages 292--301. Springer, 2018.

\bibitem{ciregan2012multi}
D.~Ciregan, U.~Meier, and J.~Schmidhuber.
\newblock Multi-column deep neural networks for image classification.
\newblock In {\em 2012 IEEE conference on computer vision and pattern
  recognition}, pages 3642--3649. IEEE, 2012.

\bibitem{dou2018peritumoral}
T.~H. Dou, T.~P. Coroller, J.~J. van Griethuysen, R.~H. Mak, and H.~J. Aerts.
\newblock Peritumoral radiomics features predict distant metastasis in locally
  advanced nsclc.
\newblock {\em PloS one}, 13(11):e0206108, 2018.

\bibitem{he2016deep}
K.~He, X.~Zhang, S.~Ren, and J.~Sun.
\newblock Deep residual learning for image recognition.
\newblock In {\em Proceedings of the IEEE conference on computer vision and
  pattern recognition}, pages 770--778, 2016.

\bibitem{he2020multi}
T.~He, J.~Hu, Y.~Song, J.~Guo, and Z.~Yi.
\newblock Multi-task learning for the segmentation of organs at risk with label
  dependence.
\newblock {\em Medical Image Analysis}, 61:101666, 2020.

\bibitem{heath2000pkegelmeyer}
M.~Heath, K.~Bowyer, D.~Kopans, and R.~Moore~Jr.
\newblock Pkegelmeyer. the digital database for screening mammography.
\newblock In {\em The Proceedings of the 5th International Workshop on Digital
  Mammography. Madison, WI, USA: Medical Physics Publishing}, 2000.

\bibitem{hu2020assessment}
Y.~Hu, C.~Xie, H.~Yang, J.~W. Ho, J.~Wen, L.~Han, K.~W. Chiu, J.~Fu, and
  V.~Vardhanabhuti.
\newblock Assessment of intratumoral and peritumoral computed tomography
  radiomics for predicting pathological complete response to neoadjuvant
  chemoradiation in patients with esophageal squamous cell carcinoma.
\newblock {\em JAMA network open}, 3(9):e2015927--e2015927, 2020.

\bibitem{huang2017densely}
G.~Huang, Z.~Liu, L.~Van Der~Maaten, and K.~Q. Weinberger.
\newblock Densely connected convolutional networks.
\newblock In {\em Proceedings of the IEEE conference on computer vision and
  pattern recognition}, pages 4700--4708, 2017.

\bibitem{janowczyk2016deep}
A.~Janowczyk and A.~Madabhushi.
\newblock Deep learning for digital pathology image analysis: A comprehensive
  tutorial with selected use cases.
\newblock {\em Journal of pathology informatics}, 7, 2016.

\bibitem{kamnitsas2017efficient}
K.~Kamnitsas, C.~Ledig, V.~F. Newcombe, J.~P. Simpson, A.~D. Kane, D.~K. Menon,
  D.~Rueckert, and B.~Glocker.
\newblock Efficient multi-scale 3d cnn with fully connected crf for accurate
  brain lesion segmentation.
\newblock {\em Medical image analysis}, 36:61--78, 2017.

\bibitem{kingma2014adam}
D.~P. Kingma and J.~Ba.
\newblock Adam: A method for stochastic optimization.
\newblock {\em arXiv preprint arXiv:1412.6980}, 2014.

\bibitem{krizhevsky2017imagenet}
A.~Krizhevsky, I.~Sutskever, and G.~E. Hinton.
\newblock Imagenet classification with deep convolutional neural networks.
\newblock {\em Communications of the ACM}, 60(6):84--90, 2017.

\bibitem{kumar2012radiomics}
V.~Kumar, Y.~Gu, S.~Basu, A.~Berglund, S.~A. Eschrich, M.~B. Schabath,
  K.~Forster, H.~J. Aerts, A.~Dekker, D.~Fenstermacher, et~al.
\newblock Radiomics: the process and the challenges.
\newblock {\em Magnetic resonance imaging}, 30(9):1234--1248, 2012.

\bibitem{lambin2012radiomics}
P.~Lambin, E.~Rios-Velazquez, R.~Leijenaar, S.~Carvalho, R.~G. Van~Stiphout,
  P.~Granton, C.~M. Zegers, R.~Gillies, R.~Boellard, A.~Dekker, et~al.
\newblock Radiomics: extracting more information from medical images using
  advanced feature analysis.
\newblock {\em European journal of cancer}, 48(4):441--446, 2012.

\bibitem{long2015fully}
J.~Long, E.~Shelhamer, and T.~Darrell.
\newblock Fully convolutional networks for semantic segmentation.
\newblock In {\em Proceedings of the IEEE conference on computer vision and
  pattern recognition}, pages 3431--3440, 2015.

\bibitem{milletari2016v}
F.~Milletari, N.~Navab, and S.-A. Ahmadi.
\newblock V-net: Fully convolutional neural networks for volumetric medical
  image segmentation.
\newblock In {\em 2016 fourth international conference on 3D vision (3DV)},
  pages 565--571. IEEE, 2016.

\bibitem{pan2009survey}
S.~J. Pan and Q.~Yang.
\newblock A survey on transfer learning.
\newblock {\em IEEE Transactions on knowledge and data engineering},
  22(10):1345--1359, 2009.

\bibitem{playout2018multitask}
C.~Playout, R.~Duval, and F.~Cheriet.
\newblock A multitask learning architecture for simultaneous segmentation of
  bright and red lesions in fundus images.
\newblock In {\em International Conference on Medical Image Computing and
  Computer-Assisted Intervention}, pages 101--108. Springer, 2018.

\bibitem{prasanna2017radiomic}
P.~Prasanna, J.~Patel, S.~Partovi, A.~Madabhushi, and P.~Tiwari.
\newblock Radiomic features from the peritumoral brain parenchyma on
  treatment-naive multi-parametric mr imaging predict long versus short-term
  survival in glioblastoma multiforme: preliminary findings.
\newblock {\em European radiology}, 27(10):4188--4197, 2017.

\bibitem{qu2019joint}
H.~Qu, G.~Riedlinger, P.~Wu, Q.~Huang, J.~Yi, S.~De, and D.~Metaxas.
\newblock Joint segmentation and fine-grained classification of nuclei in
  histopathology images.
\newblock In {\em 2019 IEEE 16th International Symposium on Biomedical Imaging
  (ISBI 2019)}, pages 900--904. IEEE, 2019.

\bibitem{ren2016faster}
S.~Ren, K.~He, R.~Girshick, and J.~Sun.
\newblock Faster r-cnn: Towards real-time object detection with region proposal
  networks.
\newblock {\em IEEE transactions on pattern analysis and machine intelligence},
  39(6):1137--1149, 2016.

\bibitem{ronneberger2015u}
O.~Ronneberger, P.~Fischer, and T.~Brox.
\newblock U-net: Convolutional networks for biomedical image segmentation.
\newblock In {\em International Conference on Medical image computing and
  computer-assisted intervention}, pages 234--241. Springer, 2015.

\bibitem{ruder2017overview}
S.~Ruder.
\newblock An overview of multi-task learning in deep neural networks.
\newblock {\em arXiv preprint arXiv:1706.05098}, 2017.

\bibitem{selvaraju2017grad}
R.~R. Selvaraju, M.~Cogswell, A.~Das, R.~Vedantam, D.~Parikh, and D.~Batra.
\newblock Grad-cam: Visual explanations from deep networks via gradient-based
  localization.
\newblock In {\em Proceedings of the IEEE international conference on computer
  vision}, pages 618--626, 2017.

\bibitem{simonyan2014very}
K.~Simonyan and A.~Zisserman.
\newblock Very deep convolutional networks for large-scale image recognition.
\newblock {\em arXiv preprint arXiv:1409.1556}, 2014.

\bibitem{szegedy2015going}
C.~Szegedy, W.~Liu, Y.~Jia, P.~Sermanet, S.~Reed, D.~Anguelov, D.~Erhan,
  V.~Vanhoucke, and A.~Rabinovich.
\newblock Going deeper with convolutions.
\newblock In {\em Proceedings of the IEEE conference on computer vision and
  pattern recognition}, pages 1--9, 2015.

\bibitem{szegedy2013deep}
C.~Szegedy, A.~Toshev, and D.~Erhan.
\newblock Deep neural networks for object detection.
\newblock {\em Advances in neural information processing systems},
  26:2553--2561, 2013.

\bibitem{szegedy2016rethinking}
C.~Szegedy, V.~Vanhoucke, S.~Ioffe, J.~Shlens, and Z.~Wojna.
\newblock Rethinking the inception architecture for computer vision.
\newblock In {\em Proceedings of the IEEE conference on computer vision and
  pattern recognition}, pages 2818--2826, 2016.

\bibitem{thome2019multitask}
N.~Thome, S.~Bernard, V.~Bismuth, F.~Patoureaux, et~al.
\newblock Multitask classification and segmentation for cancer diagnosis in
  mammography.
\newblock In {\em International Conference on Medical Imaging with Deep
  Learning--Extended Abstract Track}, 2019.

\bibitem{trullo2017joint}
R.~Trullo, C.~Petitjean, D.~Nie, D.~Shen, and S.~Ruan.
\newblock Joint segmentation of multiple thoracic organs in ct images with two
  collaborative deep architectures.
\newblock In {\em Deep Learning in Medical Image Analysis and Multimodal
  Learning for Clinical Decision Support}, pages 21--29. Springer, 2017.

\bibitem{vesal2018multi}
S.~Vesal, S.~M. Patil, N.~Ravikumar, and A.~K. Maier.
\newblock A multi-task framework for skin lesion detection and segmentation.
\newblock In {\em OR 2.0 Context-Aware Operating Theaters, Computer Assisted
  Robotic Endoscopy, Clinical Image-Based Procedures, and Skin Image Analysis},
  pages 285--293. Springer, 2018.

\bibitem{vesal2018skinnet}
S.~Vesal, N.~Ravikumar, and A.~Maier.
\newblock Skinnet: A deep learning framework for skin lesion segmentation.
\newblock In {\em 2018 IEEE Nuclear Science Symposium and Medical Imaging
  Conference Proceedings (NSS/MIC)}, pages 1--3. IEEE, 2018.

\bibitem{wang2018simultaneous}
P.~Wang, V.~M. Patel, and I.~Hacihaliloglu.
\newblock Simultaneous segmentation and classification of bone surfaces from
  ultrasound using a multi-feature guided cnn.
\newblock In {\em International conference on medical image computing and
  computer-assisted intervention}, pages 134--142. Springer, 2018.

\bibitem{yang2017novel}
X.~Yang, Z.~Zeng, S.~Y. Yeo, C.~Tan, H.~L. Tey, and Y.~Su.
\newblock A novel multi-task deep learning model for skin lesion segmentation
  and classification.
\newblock {\em arXiv preprint arXiv:1703.01025}, 2017.

\bibitem{zhang2017survey}
Y.~Zhang and Q.~Yang.
\newblock A survey on multi-task learning.
\newblock {\em arXiv preprint arXiv:1707.08114}, 2017.

\bibitem{zhou2020multi}
Y.~Zhou, H.~Chen, Y.~Li, Q.~Liu, X.~Xu, S.~Wang, D.~Shen, and P.-T. Yap.
\newblock Multi-task learning for segmentation and classification of tumors in
  3d automated breast ultrasound images.
\newblock {\em Medical Image Analysis}, page 101918, 2020.

\bibitem{zhou2021multi}
Y.~Zhou, H.~Chen, Y.~Li, Q.~Liu, X.~Xu, S.~Wang, P.-T. Yap, and D.~Shen.
\newblock Multi-task learning for segmentation and classification of tumors in
  3d automated breast ultrasound images.
\newblock {\em Medical Image Analysis}, 70:101918, 2021.

\end{thebibliography}
\end{document}